\documentclass[twocolumn,amsmath,amssymb]{revtex4}
\usepackage[english]{babel}
\usepackage[dvips]{graphicx}
\usepackage{dcolumn}
\usepackage{bm}
\begin{document}
\title{Observation and characterization of laser-driven\\Phase Space Electron Holes}
\author{\textbf{G. Sarri$^1$, M. E. Dieckmann$^2$, C. R. D. Brown$^3$, C. A. Cecchetti$^1$, D. J. Hoarty$^3$, S. F. James$^3$, R. Jung$^4$, I. Kourakis$^1$, H. Schamel$^5$, O. Willi$^4$ \emph{and} M. Borghesi$^1$}}
\affiliation{$^1$School of Mathematics and Physics, The Queen's University of Belfast, Belfast, BT7 1NN, UK\\
$^2$ITN, Linkoping University, 60174 Norrkoping, Sweden\\
$^3$AWE, Aldermaston, Reading, Berkshire RG7 4PR, UK\\
$^4$Institute for Laser and Plasma Physics, Heinrich-Heine-University, D\"{u}sseldorf, Germany\\
$^5$Physikalisches Institut, Universit\"{a}t Bayreuth, D-95440 Bayreuth, Germany}
\begin{abstract}
The direct observation and full characterization of a Phase Space Electron Hole (EH) generated by laser-matter interaction is presented. This structure has been detected via proton radiography during the interaction between an intense laser pulse ($\tau\approx$1ns temporally flat-top, I$\approx10^{14}$ W/cm$^2$) and a gold 26$\mu$m thick hohlraum. This technique has allowed us the simultaneous detection of propagation velocity, potential and electron density spatial profile across the EH with fine spatial and temporal resolution providing an unprecedentedly detailed experimental characterization.
\end{abstract}
\maketitle
Phase-Space Electron Holes (EH) \cite{schamel} are electrostatic excitations in collisionless plasmas characterized by a positive potential hump in which a population of electrons is trapped. In addition to their relevance to many fundamental plasma processes such as two-stream instabilities \cite{dieckmann} and saturation in Landau damping \cite{landau damping}, EH play a key role in a wide range of  space plasma scenarios (e.g. microscopic dissipation during magnetic reconnection in the Earth's magnetosphere \cite{drake} or the generation of cosmic ray electrons in supernovae \cite{supernovae}) and are commonly detected in near Earth plasmas \cite{mozer,hoshino}. The detrimental effect they have on the focusing properties of particle accelerators and storage rings has also been recently highlighted \cite{blask}. This omnipresence of such structures in collisionless plasmas requires therefore, beside a deep theoretical study \cite{schamel}, an equally profound experimental characterization. Previous experimental work detected this type of structures in magnetized collisionless plasmas during either strong discharge in gas tubes \cite{saeki} or magnetic reconnection in toroidal plasma current sheets \cite{fox}. In all these cases the existence of EH was deduced from positive spikes in high-bandwidth Langmuir probes. Remarkably advancing these previous detections, here we report the first direct observation of an EH in a laser-matter interaction experiment, and suggest a new way to generate and study them in a controllable manner. EH diagnosis using proton radiography \cite{sarri,lorenzo} allows in fact the simultaneous measurement of propagation velocity, potential and electron density spatial profile and temporal evolution leading to a comprehensive characterization with fine temporal and spatial resolution, of the order of few picoseconds and few microns respectively.\\
\begin{figure}[!h]
\begin{center}
\includegraphics[width=6cm,height=5.5cm]{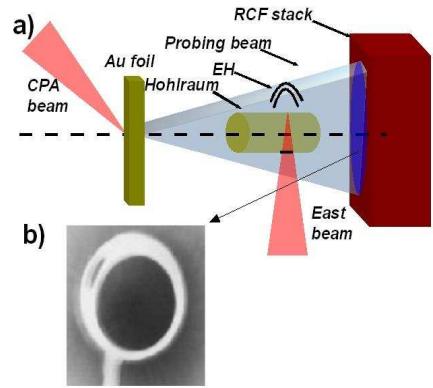}
\caption{\textbf{a. Experimental setup:} The experiment, carried out in the Helen facility \cite{AWE}, used two different laser beams.
The EAST beam ($I\approx10^{14}$W/cm$^2$, $\tau\approx1$ns flat-top, $\lambda\approx0.527\mu$m) irradiated the inner surface of a 1.5mm
diameter, $26\mu$m-thick open-ended gold hohlraum. The interaction was probed by the proton beam generated in the interaction between a $20\mu$m
gold foil and the CPA beam ($I\geq10^{19}$W/cm$^2$, $\tau\approx700$fs). \textbf{b.Proton Radiography of unirradiated hohlraum:} the Laser Entrance
Hole (LEH) is visible at the upper-left side of the hohlraum.}\label{physics}
\end{center}
\end{figure}
\verb|  |The present experiment was performed at the Helen laser system in AWE \cite{AWE}.
It involved the illumination of the inner surface of a hohlraum target by an intense and relatively long ($\tau\approx$1ns temporally flat-top, I$\approx10^{14}$ W/cm$^{2}$, $\lambda$=0.527nm) laser pulse, an interaction which produces conditions relevant to Indirect Drive Inertial Confinement Fusion (ICF) experiments \cite{lindl}. The hohlraum target consisted of an open-ended Au cylinder with diameter of 1.5 mm, length of 1 mm, and wall thickness of 26 $\mu$m. The interaction was probed via proton radiography \cite{sarri,lorenzo} and a brief sketch of the experimental setup is visible in Figure \ref{physics}. The interaction beam (EAST  beam  in Figure 1.a) was focused through a 350$\mu$m diameter entrance hole onto the inner surface of the hohlraum. 
\begin{figure*}[!t]
\begin{center}
\includegraphics[width=16cm,height=4.5cm]{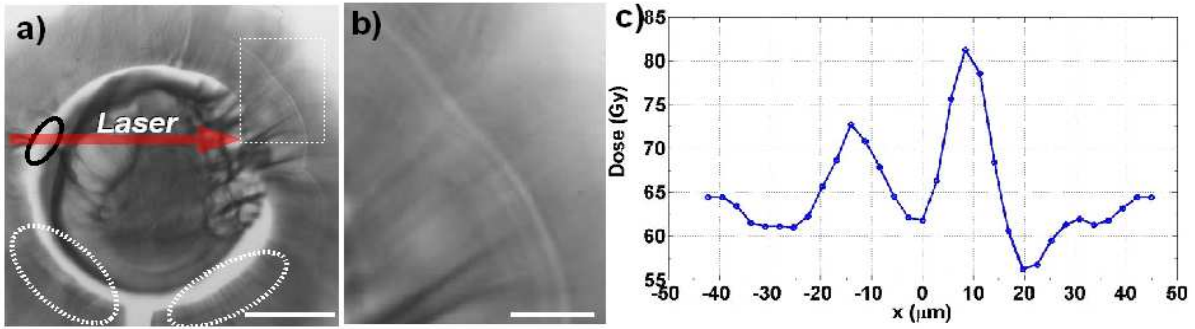}
\caption{\textbf{a. Experimental data:} RCF image associated to a time $\simeq160$ps after the beginning of interaction. The hohlraum is open-ended and has a diameter of $1500\mu$m. The dashed square in the top-right corner highlights the feature discussed in this Letter. The white ellipses outline zones of proton density depletion due to the electrostatic charging of the hohlraum walls. The white ruler represents $500\mu$m. \textbf{b. Zoom of the structure:} Zoom of the region outlined by the dashed square in \textbf{a)}.The white ruler represents $100\mu$m. \textbf{c. Proton signal:} modulation in proton dose on the RCF layer in correspondence of the structure.}\label{shot4}
\end{center}
\end{figure*}
A second short and intense pulse ($\tau\approx700$fs, I$> 10^{19}$ W/cm${^2}$, CPA beam in Figure 1.a) was focused onto a 20$\mu$m gold foil in order to create, via Target Normal Sheath Acceleration (TNSA), \cite{snavely} a wide spectrum proton beam, with a changeable delay with respect to the EAST beam. The proton beam, after having probed the plasma, was recorded on a stack of RadioChromic Films (RCF), \cite{RCF}.
Such a probing scheme enables the monitoring of the transverse electric field distribution inside the plasma by measuring the deflection of a proton beam as it passes through it. The technique exploits the fact that, as a consequence of the high degree of laminarity of the beam, the proton source, while being physically extended, is practically equivalent to a nearly point-like virtual source located in proximity to the target \cite{borghesi}. A point projection of the probed region is thus obtained with a geometrical magnification of
$M\sim10$.
Under the assumption of small deviations (i.e. the proton trajectories do not cross), the transverse electric field distribution can be derived directly from the relative modulation of the proton density deposited on a given RCF layer \cite{sarri}:
\begin{equation}
<E_y>\approx-\frac{2\varepsilon_pM}{eLb}\int\frac{\delta n_p}{n_p}\verb| |dy\label{Etrans},
\end{equation}
where $<Ey>$ is the transverse electric field component averaged along the longitudinal dimension, $\varepsilon_p$ is the probe proton energy, $L$ is the distance between the interaction area and the detector, $b$ is the longitudinal length of the non-zero electric field region and $dn_p/n_p$ is the relative modulation of the proton density at the detector plane.\\
\verb|  |Data exemplifying the features observed by proton imaging are displayed in Figure \ref{shot4}.a and \ref{shot4}.b. As a general rule, the electric fields are directed from the regions of a lighter grey color compared to the background (reduced probe proton flux) towards the regions of darker grey color (increased flux). A region of pronounced modulation in the probe proton density (Figure \ref{shot4}.c), evidence of a modulated electric field distribution, is observed $\simeq$300-400$\mu$m from the rear surface of the irradiated target surface. A region of turbulent plasma, resulting in a chaotic deflection pattern, is visible at the interaction point even though the aim of the Letter is to study the solitary structure at the rear surface of interaction.
\begin{figure}[!h]
\begin{center}
\includegraphics[width=6cm,height=4cm]{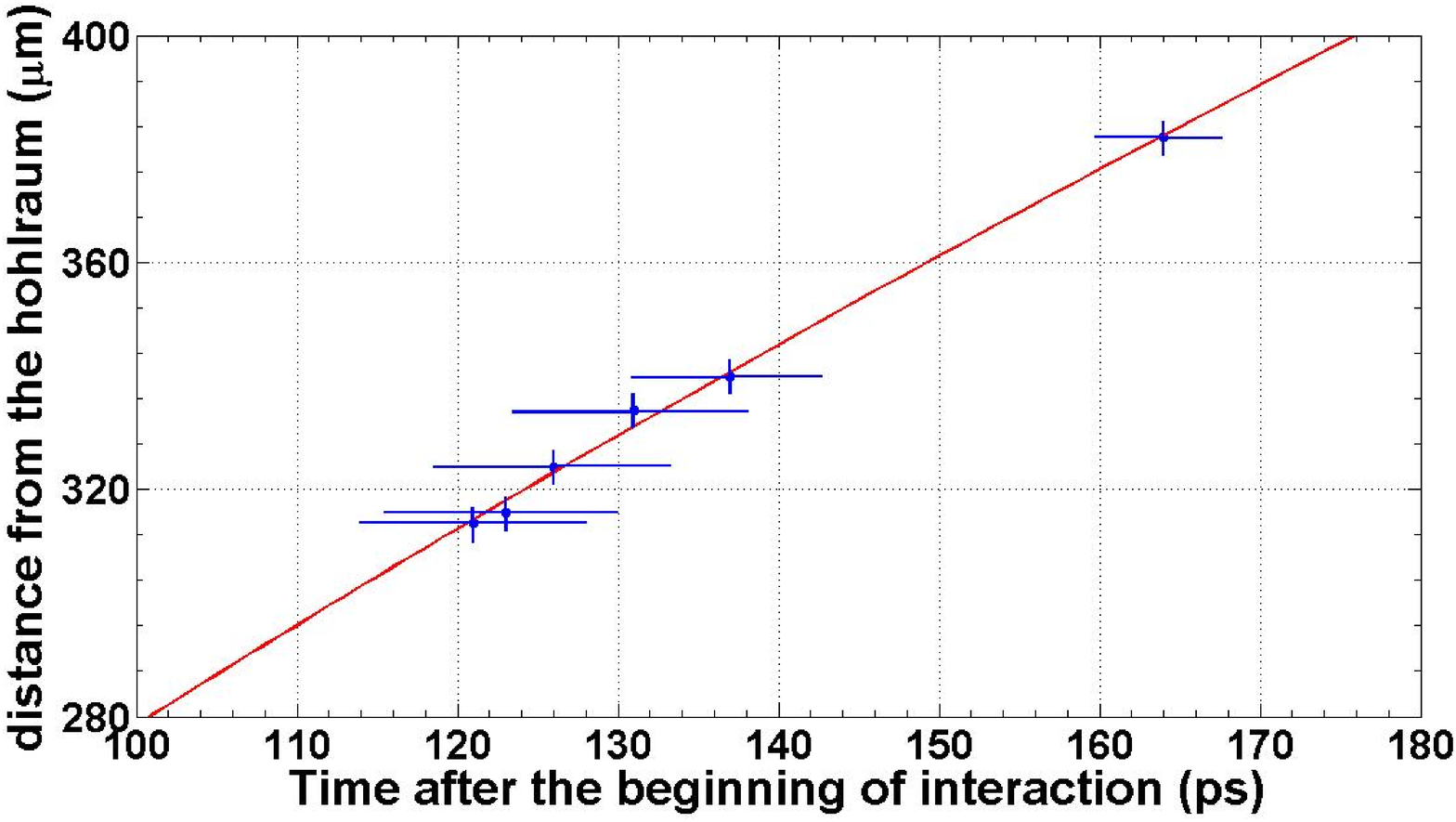}
\caption{\textbf{EH propagation} as a function of time relative to the beginning of the interaction.
The temporal error bars are due to the finite transit time of the protons across the structure.
Experimental data indicate a constant velocity of $(1.6\pm0.6)\cdot10^6$ m/s$\simeq(1.6\pm0.6)v_{th}.$}\label{propagation}
\end{center}
\end{figure}
\begin{figure}[!t]
\begin{center}
\includegraphics[width=7cm,height=13cm]{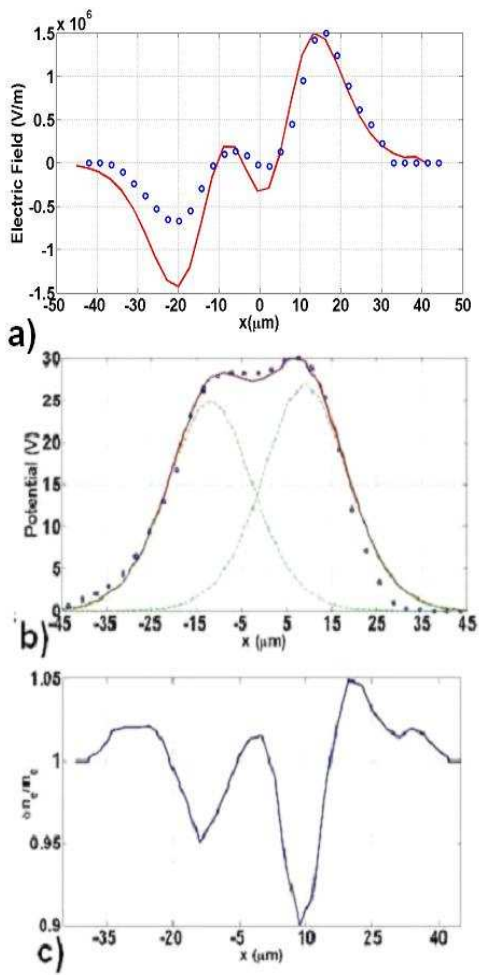}
\caption{\textbf{a) Electric field distribution:} experimental electric field distribution across the soliton structure (blue empty circles) compared with the electric field profile (red line) obtained, by using Poisson's equation, from the theoretical potential in Figure b). \textbf{b) Potential spatial profile:} experimental potential profile (blue empty circles) compared with the theoretical potential profile (red line) obtained by adding the contribution of two partially overlapping EHs (green dashed curves). \textbf{c) Electron density depletion:} experimental electron density distribution inside the EHs; two different regions of electron depletion are clearly visible suggesting the presence of two different EH structures.}\label{analisi}
\end{center}
\end{figure}
Thanks to the broad spectrum of such a proton beam, different layers within the RCF stack provide snapshots of the interaction at different times even in a single shot configuration, \cite{sarri}. This density modulation (shown in Figure \ref{shot4}.c) shows to propagate with a constant
velocity of $v\simeq(1.6\pm0.6)\cdot10^6$m/s (Figure \ref{propagation}) while maintaining a substantially time-independent profile in the co-moving
reference frame. The electric field distribution across the structure, $E(x)$, has been extracted (Figure \ref{analisi}.a) using Equation \ref{Etrans}. The corresponding potential profile has been calculated by spatial integration of $E(x)$ (Figure \ref{analisi}.b). In this calculation we have assumed a quasi-planar structure with longitudinal dimension $b$ of the order of the transverse dimension, i.e. $\approx600\mu$m; these symmetry
considerations concur with published numerical results \cite{califano}. The potential exhibits a localized bell-shaped structure $80-90\mu$m wide
with a maximum value of $\approx30$V.\\
\verb|  |In order to estimate the plasma parameters in the region of observation, the interaction between the laser and a $26\mu$m thick gold foil has been simulated using a 1D hydrodynamic Lagrangian code (Hyades) including radiation transport and ionisation \cite{haydes}.
 Simulations indicate that the more energetic X-rays generated during the interaction propagate through the gold foil and ionize the ambient gas at the rear surface (pressure $\simeq10^{-3}$ mbar) creating a steady plasma in thermal equilibrium; the electron temperature and density are predicted to be $n_e\simeq2.5\cdot10^{12}$ cm$^{-3}$ and $T_e\simeq$ 2eV respectively, implying a Debye length of $\lambda_D\simeq 7\mu$m, an electron plasma frequency of $\omega_{pe}\simeq10^{11}$s$^{-1}$ (ion plasma frequency $\omega_{pi}\simeq5\cdot10^{8}$s$^{-1}$), an electron thermal velocity of $v_{th}\simeq10^6$ m/s and an ion acoustic velocity of $c_s\simeq3\cdot10^3$ m/s. The velocity and the width of the structure are then $\simeq1.6v_{th}$ (or $\simeq530c_s$) and $\simeq10-12\lambda_D$ respectively, while the normalized maximum value of the potential is $\phi=eV/K_BT_e\simeq$15. According to these parameters, the ratio of the electron mean free path to the electron Debye length is approximately $3\cdot10^3$. The plasma is then collisionless and it can thus support propagating EHs \cite{schamel,califano}.
Since the plasma is probed at $\approx100-200$ps after the beginning of interaction and the ion plasma frequency is $\omega_{pi}\simeq5\cdot10^{8}$s$^{-1}$ ($\omega_{pi}^{-1}\approx2$ns), it is reasonable to neglect motion of the ions and to consider them as a still, neutralizing background in the plasma. Under this assumption, the electron density across the structure can be extracted from the data; the charge density, obtainable by Poisson's equation, is equal to $\rho(x)=e\cdot(n_i(x)-n_e(x))$ where $n_e(x)$ ($n_i(x)$) is the electron (ion) density distribution. Simulations indicate an ionization state $Z=1$ therefore the ion density within the structure can be expressed as: $n_i(x)=n_{i0}=n_{e0}$ leading to $n_e(x)=n_{i0}-\rho(x)/e$, (Figure \ref{analisi}.c).
The electron density exhibits two pronounced depleted regions suggesting the simultaneous presence of two partially overlapping EHs within the structure. The electron depletion is of the order of $\approx$5-10\verb|%|, well within the range of density depletions seen in published simulations ranging from 1-2\verb|%| \cite{califano} up to 15-20\verb|%| \cite{eliasson}.\\
\verb|  |For EHs in unmagnetized plasmas, the potential profile can be described in a first approximation by its small amplitude expression:
\begin{equation}
\phi(x)=\phi_{max} sech^4\left(\frac{x}{4\sqrt{\gamma_e}}\right),\label{Vteor}
\end{equation}
where $\gamma_e$ is a numerical factor ranging from 0 to 1 depending on the electron velocity distribution in the plasma \cite{goldman}.
$\gamma_e=$1 corresponds to a pure Maxwellian distribution while a smaller value of $\gamma_e$ implies a bigger deviation from the pure Maxwellian behavior or, equivalently, a bigger value of $\kappa$ in the $\kappa$-distribution \cite{summers}.
The experimental potential shape (Figure \ref{analisi}.b) can then be interpreted as the sum of two different bell potentials with $\gamma_e\approx0.7$ corresponding to $\kappa\approx4$; a strong deviation from a pure Maxwellian distribution can reasonably be due to the hot-electron population created during the laser interaction with the hohlraum walls. The comparison between the experimental and theoretical profile is shown in Figure \ref{analisi}.b where the two single potential profiles are also plotted. Differentiating this theoretical potential profile leads to a theoretical electric field distribution that reproduces the behavior of the experimental one (Figure \ref{analisi}.a). This interpretation is corroborated by simulations reported in literature \cite{califano} showing indeed the creation of 'trains' of EHs when an electrostatic perturbation is applied at the edge of the simulation box. The maximum value of the velocity distribution of the electrons trapped by the EH can be roughly estimated by equating the maximum value of the potential energy with the electron kinetic energy: $v_{te}=(2eV_{max}/m_e)^{1/2}\simeq6.7\cdot10^6$m/s$\simeq6.7v_{th}$. The velocity oscillations within the electrostatic potential exceed the observed propagation speed $(1.6\pm0.6)\cdot10^6$m/s of the structure by a factor 4, implying that this potential must trap the electrons of the ambient plasma, which is at rest. The velocity oscillations of the trapped electrons exceed the simulated thermal speed by a factor $6-7$, which is in accordance with theory \cite{schamel} and also not unusual for an EH in simulations \cite{dieckmann}.\\
\verb|  |The perturbation driving the EH generation is thought to be the sudden charging of the hohlraum walls due to the residual positive charge in the target left behind by the accelerated electrons that are energetic enough to escape \cite{borghesi2,quinn}.
 This potential can be estimated from the deflection of the protons passing close to the walls (highlighted in Figure \ref{shot4}.a). Contiguous regions of proton depletion and accumulation are in fact present in proximity of the walls consistent with an electrostatic potential at the wall surface of $\approx800$V $=400K_BT_e/e$.
 A significant difference in amplitude between the EH and the driving potential is reported both in experiments \cite{guio} and simulations \cite{califano} and it thus appears to be a necessary condition for the excitation of such structures.\\
\verb| |The experimental propagation velocity of $(1.6\pm0.6) v_{th}$ slightly exceeds the allowed velocity range $v_{EH}\leq1.307v_{th}$ set by the analytical theory developed by Schamel \emph{et al.} \cite{schamel}.
However, it has to be noted that this range is valid only for a pure Maxwellian distribution and for EHs that have reached stationarity. The significant deviation from a Maxwellian distribution that the data suggest, together with the very early times at which the structure is observed ($t\cdot\omega_{pe}\approx16$), could be a possible explanation for this partial disagreement.
Indeed, simulations by Califano \emph{et al.} \cite{califano} show propagation velocities ranging from a fraction of $v_{th}$ up to $2v_{th}$.
The extremely high ion Mach number of the structure ( M=530) definitely excludes the possibility of an ion wave driven phenomenon such as a collisionless shock \cite{lorenzo2}.\\
\verb|  |Published analytical models \cite{schamel} and numerical simulations \cite{califano,eliasson} set the conditions to identify a solitary structure in a plasma as an EH to be: a positive hump-like potential, a depletion in the electron density at the potential peak and finally a propagation velocity of the order of the thermal electron velocity. The results shown in Figures \ref{propagation} and \ref{analisi} satisfy all the requirements expressed above. Even the spatial width of $\approx10-12\lambda_D$ concurs with the hypothesis of an EH detection; simulations \cite{califano} and space plasma measurements \cite{hoshino} show that these structures have a typical width ranging from a few up to tens of Debye lengths.\\
\verb|  |In summary, we have reported the first laboratory observation of EHs in a laser-plasma experiment. Thanks to its fine temporal and spatial resolution, of the order of few picoseconds and few microns respectively, proton radiography allows for a comprehensive characterization of the structure. Data analysis indicates the generation of two partially overlapping EHs moving with a velocity close to the thermal electron velocity. Electric field, potential profile and electron density distributions have been measured, together with the propagation velocity and width of the structure, giving results in good agreement with published analytical theory and simulations. Temporal evolution of the structure has also been accurately measured. The experimental setup shown permits to create and comprehensively characterize EHs, or other soliton-like structures, that have many implications in a wide range of plasma scenarios. Such structures are one of the most widespread nonlinear phenomena in collisionless plasma ranging from Earth's vicinity and supernova scenarios to particle accelerator storage rings. By controlling the characteristics of the ambient plasma or the laser parameters, it may be possible to study this type of phenomena in a wide range of different regimes opening up the possibility of acquiring unprecedented experimental knowledge in this field.\\
\verb|  |Funding for this research has been provided by the AWE Academic Access Scheme, EPSRC grants EP/E035728/1 and EP/C003586/1, by DFG TR 18, GK 1203 and FOR 1048 VR. The authors acknowledge the support of the HELEN Facility laser and target preparation personnel and K.Quinn for the fruitful help given.\\ \\

\end{document}